\documentstyle[aps,eqsecnum,prd,epsfig,multicol]{revtex}
\begin{document}

\title{Stability of self-gravitating magnetic monopoles}

\author{Guillermo Arreaga\footnote{Electronic address:
garreaga@fis.cinvestav.mx}}
\address{Departamento de F\'{\i}sica, Centro de 
Investigaci\'on y de Estudios Avanzados del IPN\\
Apdo. Postal 14-740,07000 M\'exico, DF, MEXICO}

\author{Inyong Cho\footnote{Electronic address: 
cho@physics.emory.edu}}
\address{Department of Physics, Emory University,
Atlanta, Georgia 30322-2430, USA}

\author{Jemal Guven\footnote{Electronic address:
jemal@nuclecu.unam.mx}}
\address{Instituto de Ciencias Nucleares, Universidad Nacional Aut\'onoma 
de M\'exico\\
Apdo. Postal 70-543, 04510 M\'exico, DF, MEXICO}


\maketitle

\begin{abstract}
The stability of a spherically symmetric
self-gravitating magnetic monopole is examined 
in the thin wall approximation: modeling the interior false vacuum 
as a region 
of de Sitter space; the exterior as an asymptotically flat region of the 
Reissner-Nordstr\"om geometry; and the boundary separating 
the two as a charged domain wall. There remains only to 
determine how the wall gets embedded in these two geometries.
In this approximation, the ratio $k$ of the false vacuum to surface 
energy densities is a measure
of the symmetry breaking scale $\eta$. Solutions  are characterized by 
this ratio, the charge on the wall $Q$, 
and the value of the conserved total energy $M$. 
We find that for each fixed $k$ and $Q$ up to some
critical value, there exists a 
unique globally static solution, with  
$M\simeq Q^{3/2}$; any
stable radial excitation has $M$ bounded above  by $Q$,  the
value assumed in an extremal Reissner-Nordstr\"om geometry and 
these are the only solutions with $M<Q$. As $M$ is raised above $Q$ 
a black hole forms in the exterior: (i) for low
$Q$ or $k$, the wall is crushed; (ii) for higher values,
it oscillates inside the black hole. 
If the mass is not too high these `collapsing'
solutions co-exist with an inflating bounce; (iii) for $k$, $Q$ or $M$ 
outside the above regimes, there is a unique inflating solution. 
In case (i) the course of the bounce lies within a single 
asymptotically flat region (AFR) and it resembles closely the
bounce exhibited by a false vacuum bubble (with $Q=0$). 
In cases (ii) and (iii) the course of the bounce spans two 
consecutive AFRs. However, from the 
point of view of either region it resembles a 
monotonic false vacuum bubble.
\end{abstract}
 
\begin{multicols}{2}

\section{Introduction}
\label{int}

Several years ago, Linde and Vilenkin pointed out the
possibility that the core of a localized topological defect 
could inflate under appropriate conditions in a process that was aptly 
dubbed topological inflation \cite{Linde,Alex}. 
See also \cite{Eduardo}.

A common characteristic of such defects 
is some non-linear scalar field (the Higgs field)
forced up in its core into a constant excited false vacuum state, 
falling through a transition layer to the true vacuum value 
remote from the core.

Without further refinement, a configuration of this type  will always
collapse, with or without gravity. The dynamics of this basic
configuration, a so called false vacuum bubble,
was studied in detail in the eighties by various groups, perhaps most 
comprehensively by Blau, Guendelman and Guth (BGG) in Ref.\cite{bgg} where 
references to the earlier literature are provided.
Their work was motivated, like Linde and Vilenkin's, by the 
possibility that a false vacuum bubble could provide
a seed for an inflationary universe \cite{ag}.

They discovered that the essential radial dynamics of the scalar field 
is captured extremely accurately by a simple one dimensional 
mechanical caricature of the field configuration. This 
led them to consider the dynamics of a spherically symmetric
region of false vacuum which is separated by a domain wall 
from an infinite true vacuum exterior (with zero energy density).  

The energy momentum tensor of false vacuum is inherently homogeneous 
and isotropic. The Birkhoff theorem  then constrains the 
spacetime it occupies to coincide with some patch of de Sitter space; 
the exterior is simply the Schwarzschild geometry 
truncated at the false vacuum boundary. In this model, it becomes clear
that collapse does not necessarily spell the demise of the false 
vacuum interior. When gravity is taken into account,
there are (at least) two spherically symmetric configurations 
associated with each sufficiently low value of conserved 
Arnowitt-Deser-Misner (ADM) mass
$M$. While the smaller of the two, no matter what its initial radius,  
will always collapse to a vanishing radius, 
the other will inflate; but unexpectedly from an  
euclidean perspective, not at the expense of the exterior.
This peculiar state of affairs is possible because the expansion 
occurs behind a minimal surface (a wormhole) in the Schwarzschild 
geometry connecting the false vacuum core to the exterior. 
While false vacuum is destroyed by the motion of the boundary,
it is created exponentially faster in the interior. 
The wormhole itself, however, always collapses into a black hole. 

Field theories admitting static configurations when gravity is turned on
were first constructed in the eighties. Self-gravitating global monopoles 
were considered by Barriola and Vilenkin
\cite{bv} (with additional insight provided in
Ref.~\cite{hl}). Gauge monopoles were
considered by Gibbons \cite{gibb}, with the subsequent
numerical solution of the static Einstein-Yang-Mills-Higgs equations by 
Ortiz \cite{ortiz} and Breitenlohner, Forgacs and Maison \cite{bfm}.
However, as Gibbons himself realized the balance of 
forces is not always possible when gravity is acting. 
As Linde and Vilenkin were later to show, 
if the symmetry-breaking scale
is increased towards the Planck scale, at some point the interior radius 
will exceed the corresponding cosmological horizon, triggering the 
inflation of the interior. A bound exists beyond which static
configurations necessarily become unstable. 
This process was studied by
Sakai {\it et al.}~\cite{sstm,sk} by numerically solving 
the dynamical Einstein-Higgs and Einstein-Yang-Mills Higgs equations. 
The inflating core was not unlike
an inflating false vacuum bubble. In Ref.\cite{IJ}
two of us showed how the model of a false vacuum bubble 
could be adapted to imitate the dynamics of
a self-gravitating global monopole under these extreme conditions. 
Technically this was simple, 
involving the substitution of the Barriola-Vilenkin geometry describing the 
asymptotics of a global monopole for the exterior Schwarzschild geometry of 
the former. It was possible to capture, 
remarkably faithfully, the essential underlying physics of 
topological inflation found earlier by Sakai {\it et al.}
The onset of topological inflation 
was very clearly indicated at $8\pi \eta^2 =1$ 
(in natural units).

The details of topological inflation in a gauge monopole 
are very different. Here, the only long range field is the magnetic 
Coulomb field and the 
total energy is finite.  However, the 
source resides on the core boundary which 
suggests the same mechanical caricature: de Sitter space inside,
a domain wall, but with the Reissner-Nordstr\"om geometry outside.
The structure of this geometry is very different from Schwarzschild.
If the conserved charge to mass ratio of the configuration $Q/M\le 1$,
the analyically continued geometry possesses 
horizons, otherwise it contains a naked singularity. 
In this paper, we examine the above model in detail.
Though simple in principle, a thorough analysis of the three-dimensional
parameter space $(\eta$, $Q$, $M$) is  complicated in practice. 
We will focus on the identification of the regimes of
parameter space admitting solutions which are either static, collapsing or
inflating within the core and we will examine the fate of these 
solutions as the relevant boundaries in parameter space are crossed. 

Our results can be summarized as follows:

\begin{enumerate}  

\item Fix $\eta$: for each non-vanishing value of $Q$ 
up to some critical 
value $Q_{0}$ there exists a unique stable and globally static 
configuration with a fixed core radius (a monopole) with
mass $M_0\sim Q (Q/Q_0)^{1/2}$. Stable radial oscillations of 
this configuration exist for all $M<Q$ above $M_0$.
These are the only solutions with $M<Q$.

Now raise $M$ above $Q$ but below some value $M_+$:

\item (i) For non-vanishing values of $Q$ up to some 
value $Q_{+}$ lower than $Q_{0}$, 
this radially oscillating solution falls through an event horizon 
and is terminated by the collapse of the exterior into a 
Reissner-Nordstr\"om black hole;  
(ii) for higher (but bounded) values of $Q$  
radial oscillations lie within the inner horizon
and get isolated by the collapse of the exterior (a monopole inside 
a black-hole); an inflating bounce 
co-exists with these solutions; (iii) if $Q$ or $M$ lies outside these 
two regimes, but $M$ lying above some minimum value, there is a  unique  
inflating bounce.

\item The bounce occuring in these three regimes can be 
characterized roughly as follows:
(i) the course of the bounce lies within a single 
asymptotically flat region (AFR) and it resembles closely the
bounce exhibited by a false vacuum bubble (with $Q=0$); 
(ii),(iii) the course of the bounce spans two 
consecutive AFRs. However, from the 
point of view of either region it resembles a 
monotonic false vacuum bubble;
In all cases, the expansion takes place behind an event horizon.
These configurations are analogues in this model of the topologically 
inflating solutions observed numerically by 
Sakai \cite{sk} for large $\eta$. 

\item All collapsing and  monotonic solutions are ruled 
out as either unphysical or inconsistent with asymptotically flat
boundary conditions. 

\end{enumerate}

Aspects of the model have been examined before. 
Indeed in the sixties, it received its first incarnation in 
Dirac's proposal (without spin!) of
a model of the electron as a closed charged 
conducting membrane surrounding a vacuum interior\cite{Dirac}. 

Tachizawa, Maeda and Torii focused on the 
stability of the monopole from the point of view of catastrophe 
theory~\cite{tachi}, modeling the monopole core and exterior 
as we do but without an intermediate surface layer, a model 
originally proposed by Lee, Nair and Weinberg, in Ref.~\cite{lee}.
In this limit, the core radius exceeds the cosmological horizon 
when $\eta \sim 0.33$, signaling inflation.
However, without the domain wall to transmit energy from the false 
vacuum, all dynamical possibilities are not 
faithfully represented. More recently, Alberghi, Lowe and 
Trodden \cite{trodden} considered the model within the context 
of the Anti- de Sitter space/conformal field theory correspondence. For this 
purpose they catalogued accurately the possible trajectories of the charged
false vacuum bubble. However, they did not consider how these 
trajectories depend on the values of $Q$, $M$ or $\eta$ and they did not 
consider the parameter regime $M < Q$ corresponding to stable configurations.

The paper is organized as follows. In Sec.~\ref{model} we introduce 
the model. In 
Sec.~\ref{ssfK}, we determine all possible trajectories of the wall radius 
consistent with set values of $Q$, $\eta$ and  
$M$. In Sec.~\ref{ebwm} 
we describe briefly the interior and exterior
spacetimes and how the routing of trajectories is determined in each.
In Secs.~\ref{bwtst} - \ref{BGG},
we identify all physically interesting solutions and
compare our results with earlier work. Finally,
in Sec.~\ref{con} we conclude with a few brief comments.

\section{The model} 
\label{model}

The configuration possesses a core in which the 
magnitude of the Higgs field approximates its false vacuum value, 
$\phi=0$;
in the core region, the potential energy of 
the Higgs field dominates the gradient energy 
in the Higgs and gauge fields. We model this core by a 
spherically symmetric region of 
false vacuum, and for the Mexican sombrero potential

\begin{equation}
V(\phi) = {\lambda\over 4} (\phi^2- \eta^2)^2\,,
\end{equation}
this energy density
is given by $V(\phi=0)={\lambda \over 4} \eta^4$. The corresponding 
spacetime is then described by the de Sitter line element

\begin{equation}
ds^2 = -A_D\, dT_S^2 +{1 \over A_D}dR^2 +R^2d\Omega^2\;,
\end{equation}
where
\begin{equation}
\label{AD}
A_D=1-H^2R^2\;.
\end{equation}
The Hubble parameter $H$ appearing here
is given by $H^2={8\pi \over 3}
V(0)={2\pi\lambda \over 3} \eta^4$.

We will suppose that there is a charge $Q$ localized on the boundary of this 
core. The energy in the neighborhood of the core is dominated by field 
gradients. This boundary layer can be modeled as a relativistic domain wall 
with a surface energy density (tension) $\sigma\sim \eta^3$, \cite{CG}.

Outside the core, the energy density in the massive fields falls 
off exponentially fast so that, to a good approximation,  the 
energy in matter is dominated by the asymptotic magnetic Coulomb field.  
The spherically symmetric 
exterior spacetime can then be modeled as a region of the  
Reissner-Nordstr\"om geometry described by the line element

\begin{equation}
ds^2 = -A_M\,dT_M^2 +{1 \over A_M}dR^2 +R^2d\Omega^2\;,
\end{equation}
where

\begin{equation}
\label{AM}
A_M=1-{2M \over R}+{ Q^2 \over R^2}\;.
\end{equation}
Here $M$ is the conserved ADM mass which represents the combined
material and gravitational binding energy of the configuration.
$M$ must be positive.\footnote{ The charge 
$Q$ appearing here is related to the magnetic charge of the monopole 
$g$ by $Q^2={g^2 \over 4\pi}$ where $g={4\pi \over e}$ and $e$ is the
gauge coupling strength.}

In this model, we attempt to capture the dynamics of the 
bubble wall in a single variable, the radius $r$ of the core 
boundary or wall. Following Ref.\cite{bgg}, it is 
straightforward to cast the Einstein equations at the wall in the form 

\begin{equation}
\label{eq=dbeta}
\beta_D- \beta_M = 4\pi  \sigma r \equiv \kappa r\,,
\end{equation}
where we define $\beta_{D,M}= \pm \sqrt{\dot r^2 + A_{D,M}}$, 
and the overdot represents a derivative with respect to proper time. 
Equation~(\ref{eq=dbeta}) can be exploited to express 
both $\beta_D$ and $\beta_M$ as functions of the wall radius:

\begin{equation}
\beta_{D,M} = {1\over 2kz^3}[-(1 \mp k^2)z^4+2mz-q^2]\,,\label{eq=betadm}
\end{equation}
where we have rescaled variables as follows:

\begin{equation}
\kappa /H=k\,,\quad H M=m\,,\quad H^2Q^2=q^2\,,\quad Hr=z\quad.
\label{eq=parameter}
\end{equation}
Now Eq.~(\ref{eq=dbeta}) can be recast as 

\begin{equation}
\label{masss}
\dot{z}^2+U(z)=-1\,,
\end{equation}
where the overdot represents a derivative with respect to proper 
time rescaled by $H$.
The potential $U(z)$ appearing here
can be expressed in either of two equivalent forms

\begin{equation}
\label{eq=U}
U(z) = -\beta_D^2-z^2 =-\beta_M^2 -{2m\over z}+{q^2\over z^2}\,.
\end{equation}
The Einstein equations determine the 
local geometry in the neighborhood of the wall.
The sign of the functions $\beta_{D,M}$ encodes the
boundary conditions required to construct the complete global geometry.

Finally, we note that, in terms of 
the symmetry-breaking scale $\eta$, the ratio $k$ is given by 

\begin{equation}
\label{Keta}
k=\sqrt{24\pi\over \lambda} \, s \eta \,.
\end{equation}  
Here, we have exploited the fact that
$\rho \sim \eta^4$ and $\sigma \sim \eta^3$
with constants of proportionality  
$\lambda$ and $s$ of order unity. For a GUT symmetry-breaking
scale, $\eta\sim 10^{16}$GeV, 
$k\sim 10^{-3}$. For Planck scale $\eta$, $k\sim 1$. 

\section{All local solutions} 
\label{ssfK}

The potential $U$ given by Eq.(\ref{eq=U}) is
parametrized by three positive dimensionless 
parameters characterizing the mass, the charge, and 
the symmetry breaking scale $m$, $q$ and $k$ respectively.
We will consider sections of constant $k$ and of constant
$Q$ of this three dimensional space.
Because both $m$ and $q$ have $\eta$ folded into their 
definition, when we vary $k$, it is appropriate 
to undo the `natural' scaling depending on $\eta$  
one exploits for calculational purposes. 

\begin{figure}
\begin{center}
\epsfig{file=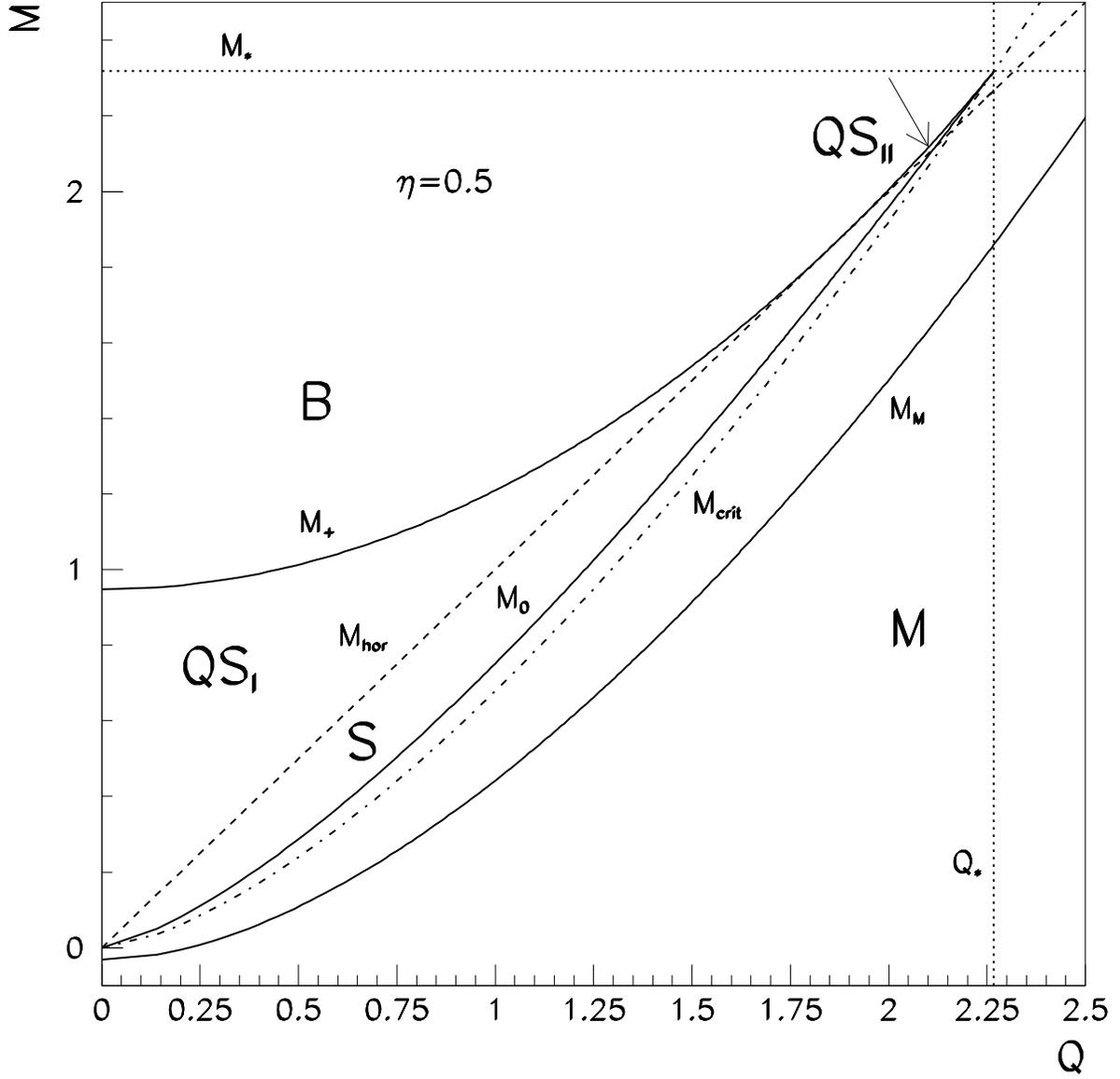}
\vspace{1 cm}
\caption{
$M-Q$ section of parameter space for $\eta=0.5$ indicating the 
following boundaries:
(i) the lower bound on the mass
providing the potential $U(z)$ with a well, $M_{crit}$;
(ii) the upper limit on monotonic motion $M_{\cal M}$;
(iii) the limits of oscillatory motion, $M_0$ and $M_+$.
$M_+$ and $M_0$ terminate at $M_*$ on $M_{crit}$.
The extremal exterior Reissner-Nordstr\"om geometry occurs at 
$M=Q$ and is indicated $M_{hor}$. The parameter regimes 
${\it S}$, ${\it QSI}$, ${\it QSII}$, ${\it B}$, and ${\it M}$ 
are indicated. For details 
see the text.}
\label{regmq}  
\end{center}
\end{figure}

\begin{figure}
\begin{center}
\epsfig{file=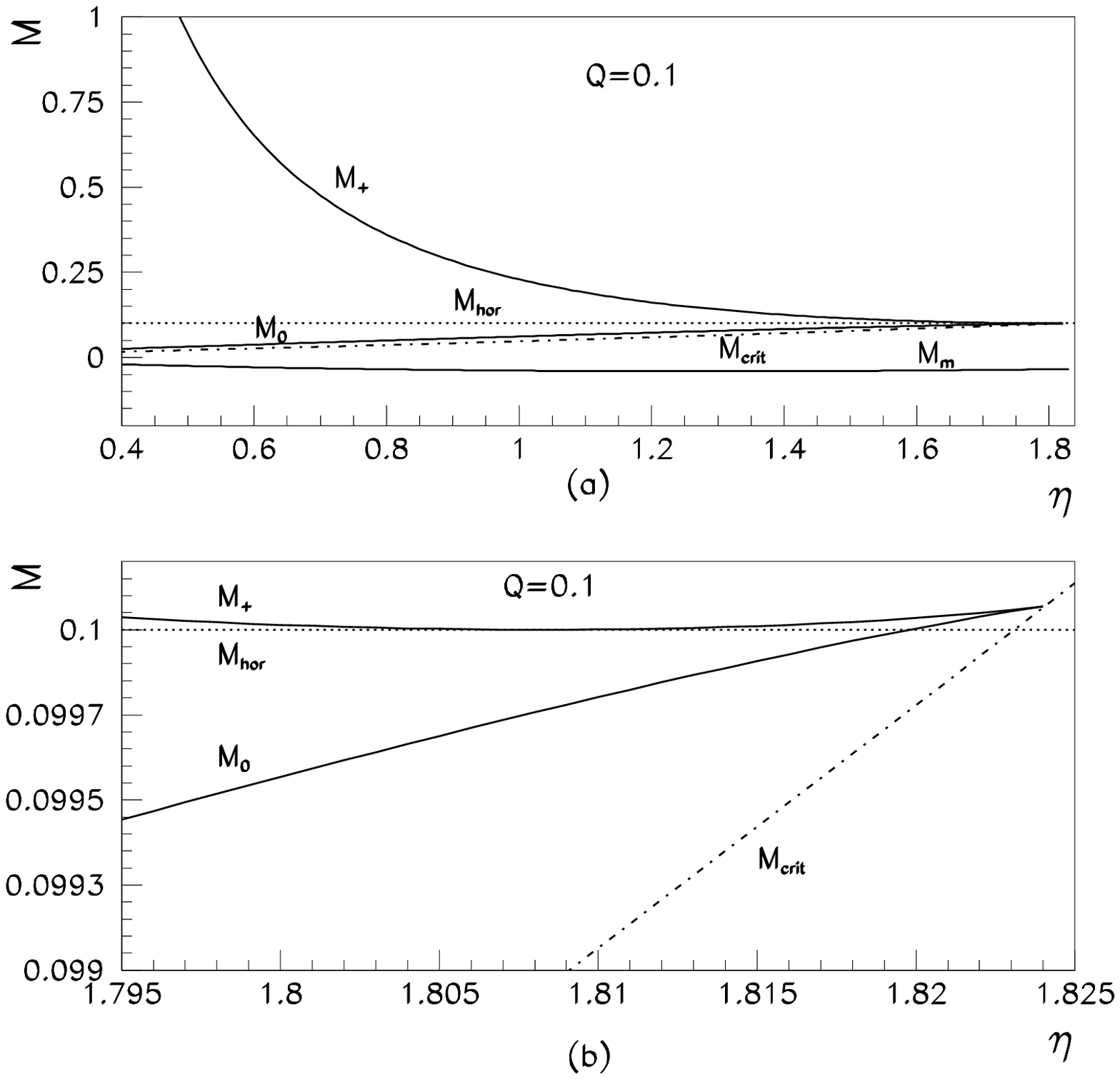}
\vspace{1 cm}
\caption{
(a) $M-\eta$ section of parameter space for $Q=0.1$ indicating the 
same boundaries as in Fig.~\ref{regmq}. 
(b) Zoom-in of the neighborhood of the bifurcation point.}
\label{regmkambos} 
\end{center}
\end{figure}

\begin{figure}
\begin{center}
\epsfig{file=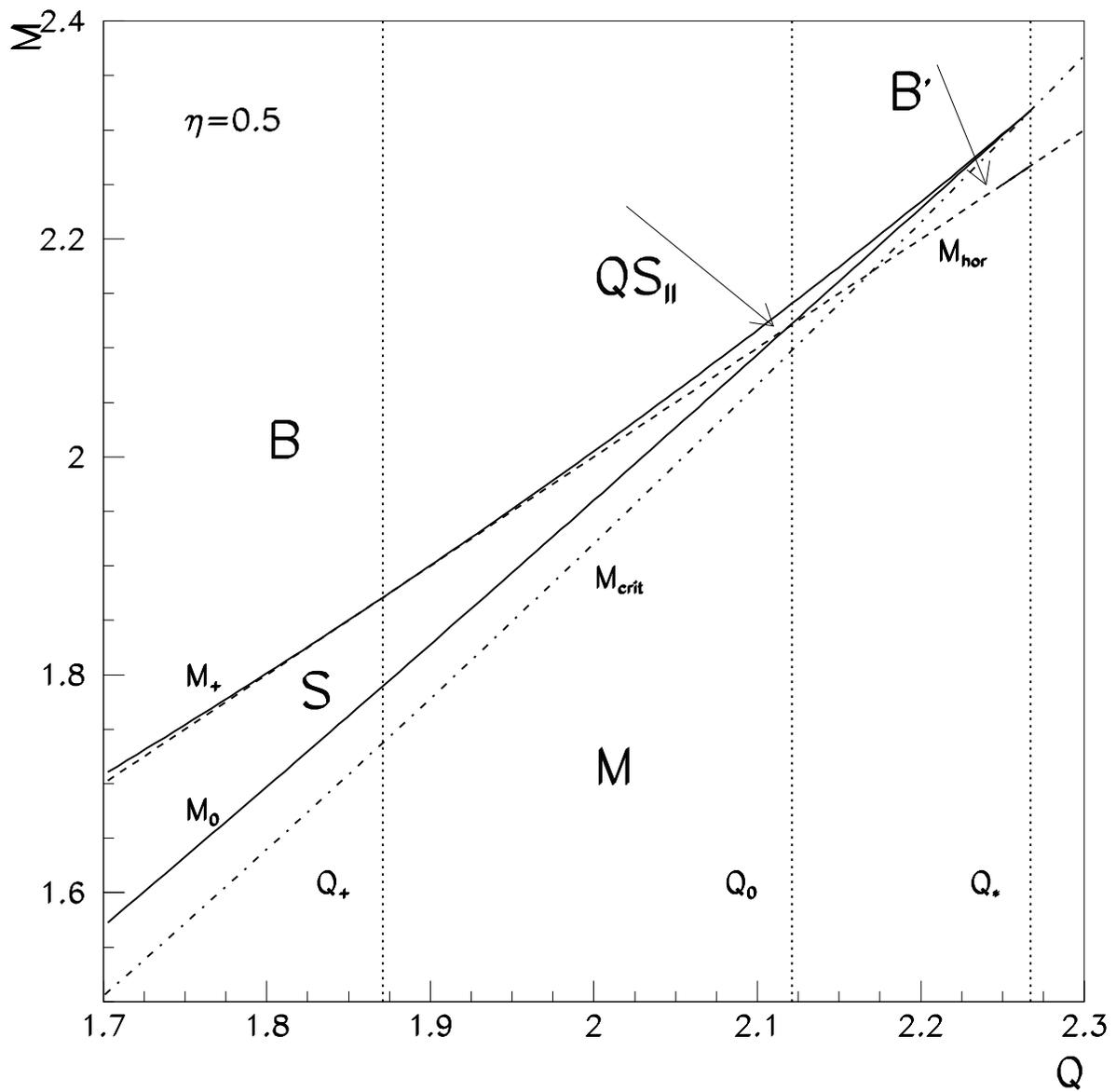}
\vspace{1 cm}
\caption{Zoom-in of the neighborhood of 
the bifurcation point $Q_*$ on Fig.~\ref{regmq}. The parameter regime
${\it B}'$ is indicated.}
\label{regmqzi}
\end{center}
\end{figure}

\begin{figure}
\begin{center}
\epsfig{file=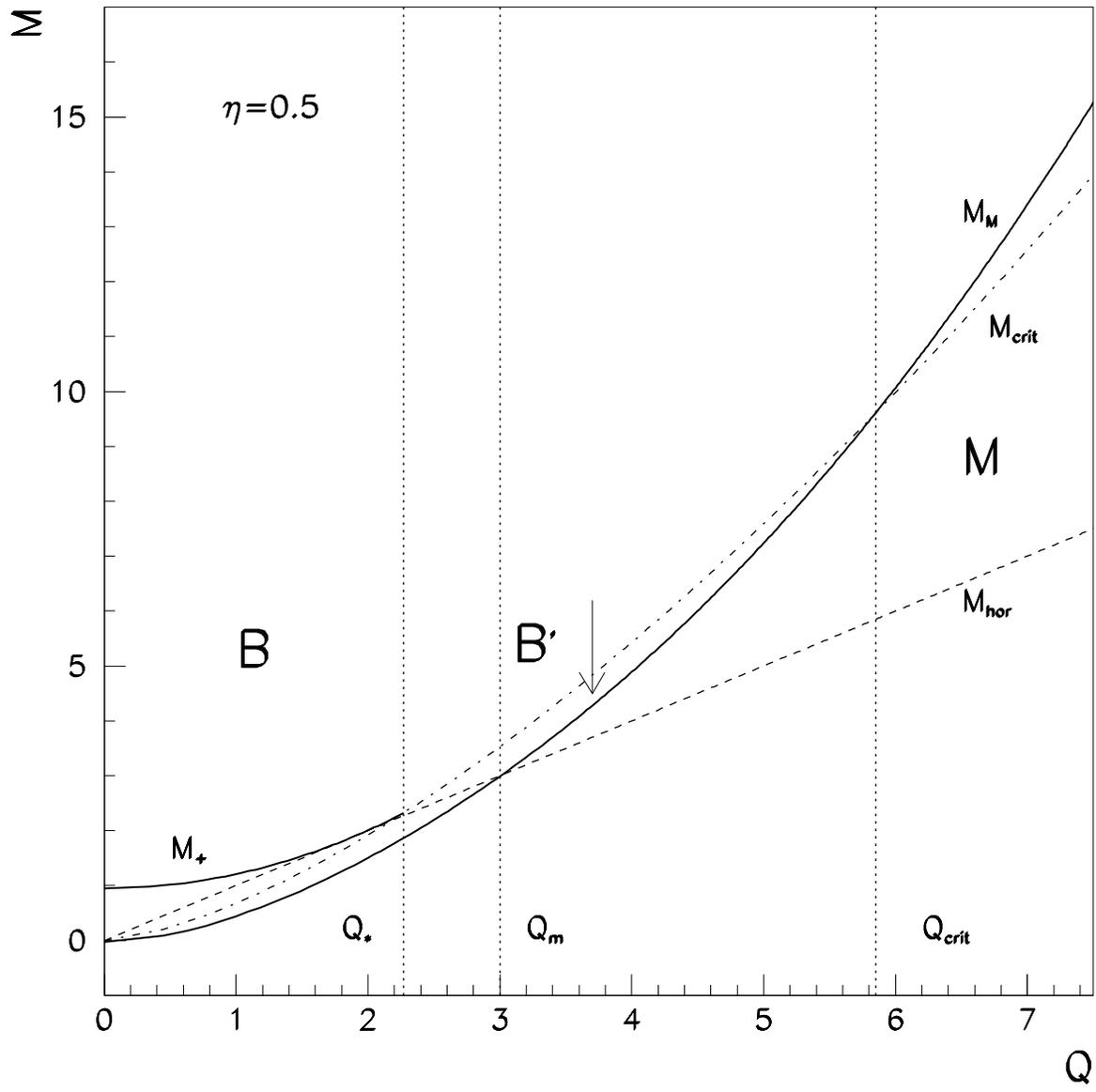}
\vspace{1 cm}
\caption{Zoom-out of Fig.~\ref{regmq}. 
For clarity, the curve $M_0$ is not indicated.}
\label{regmqzo} 
\end{center}
\end{figure}

\begin{figure}
\begin{center}
\epsfig{file=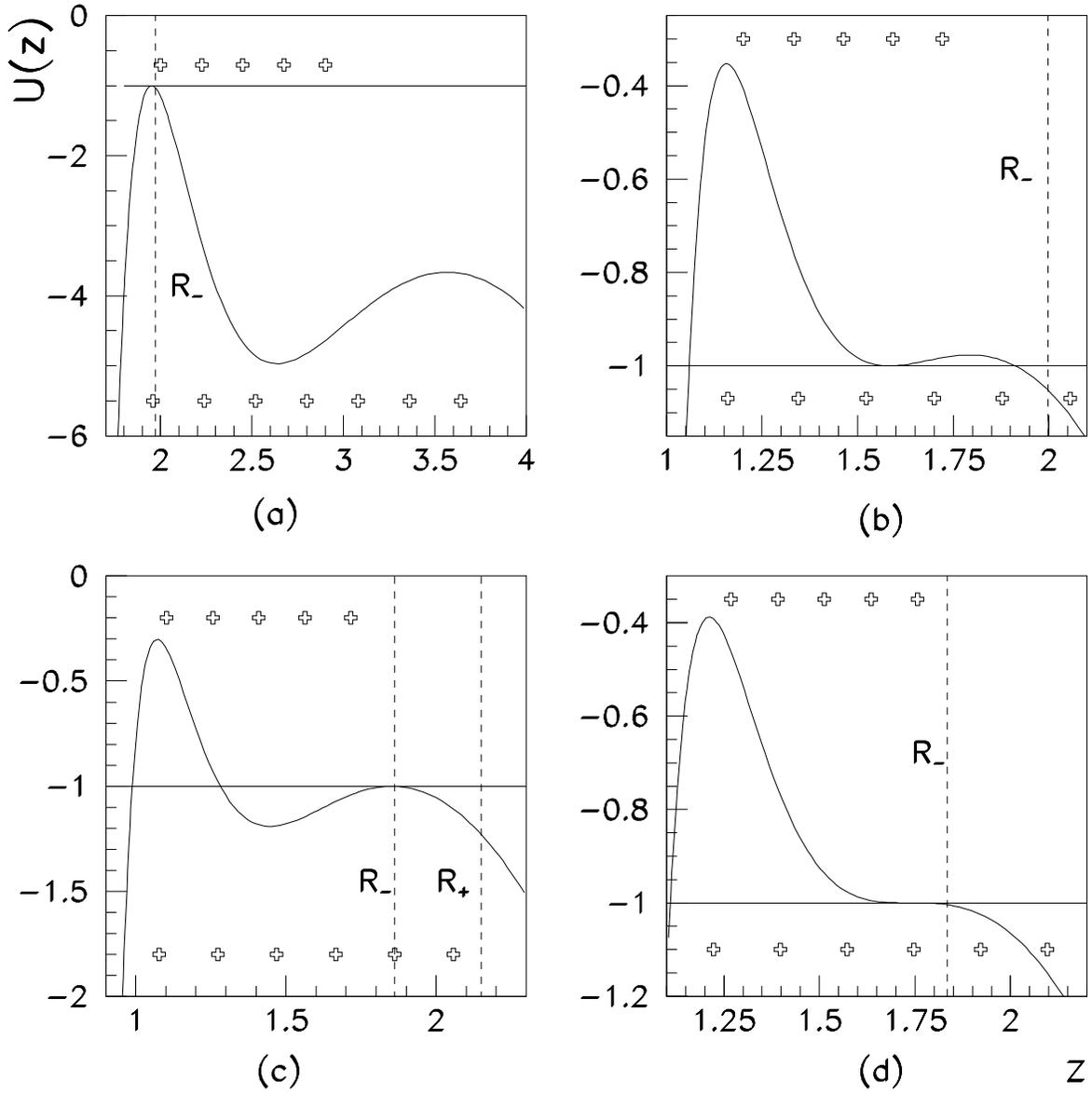}
\vspace{1 cm}
\caption{
Plot of the effective potential $U(z)$ vs. $z$ for $\eta=0.5$ 
corresponding to values of $Q$ and $M$ lying on the boundaries
$M_{\cal M}$, $M_0$, $M_+$ and at the bifurcation point, $M_*$:
(a) $M_{\cal M}$ ($Q=7.0$, $M=13.4028$);
(b) $M_0$ ($Q= 2.1307$, $M=2.1351$);
(c) $M_+$ ($Q=2.0$, $M=2.0$); 
(d) the bifurcation point $M_*$ ( $Q_*= 2.2671$, $M_*= 2.3181$).
The Reissner-Nordstr\"om horizons are located at $R_+$ and $R_-$.
The domain of positive $\beta_D$ ($\beta_M$) 
is indicated at the top (bottom) of each plot.}
\label{npot}
\end{center}
\end{figure}

\begin{figure}
\begin{center}
\epsfig{file=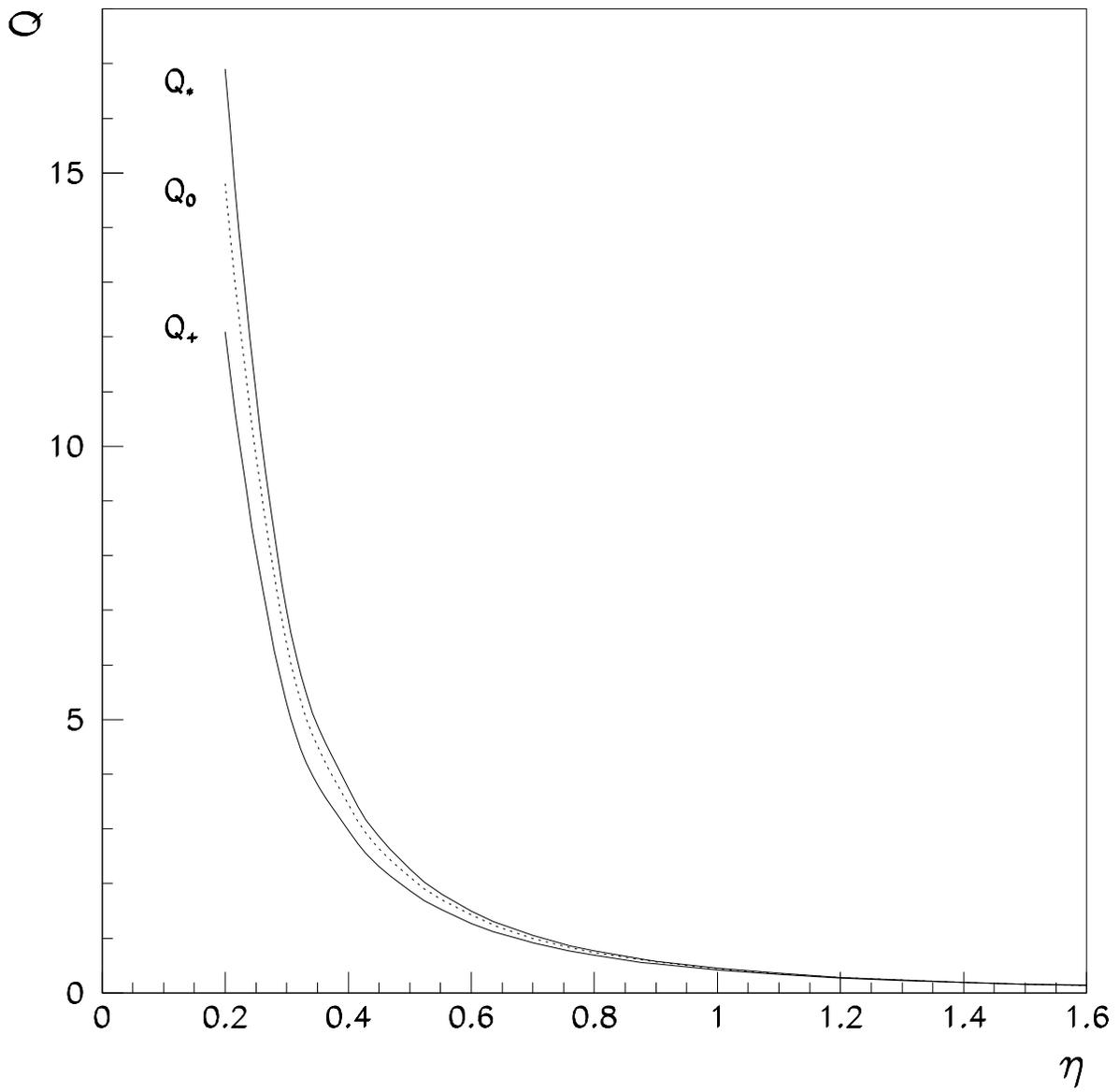}
\caption{$Q_*$, $Q_0$ and $Q_+$  vs. $\eta$}
\label{qtfinal}
\end{center}
\end{figure}

In general, the potential is always negative. In addition,
$U\to -\infty$ at $z=0$ and as $z\to\infty$ so that
it always possesses at least one maximum.
To discuss the qualitative dependence of the potential on 
the values of $M$, $Q$ and $k$, it is useful to 
identify the following boundaries on the parameter space:

\begin{enumerate}

\item The location of the extremal exterior 
Reissner-Nordstr\"om geometry, $M_{hor}$: $M=Q$

If $M>Q$ the complete Reissner-Nordstr\"om
geometry  possesses an (outer) event horizon at $R_+$ and an (inner) 
Cauchy horizon at $R_-<R_+$ which are given by the two positive solutions 
of $A_M=0$ where $A_M$ is given by Eq.(\ref{AM}).
When $M=Q$, the two horizons possess the same radius 
(this does not mean that they coalesce).
If, however, $M<Q$ there are no horizons and the corresponding spacetime
possesses a naked singularity at $R=0$. 
This criterion is independent of $k$. 

This boundary will play an important role
in determining the limit of stability 
of a self-gravitating object. 

We refer the reader to the $M-Q$ and $M-k$
parameter planes represented in Fig.~\ref{regmq} and Fig.~\ref{regmkambos}
respectively. 
As a visual aid, the former is reproduced
zoomed-in as
Fig.~\ref{regmqzi} and zoomed-out as Fig.~\ref{regmqzo}.
The corresponding potential in different regions 
of parameter space is plotted in Fig.~\ref{npot}. 

\item The lower bound on the mass
providing a potential with a well, $M_{crit}$:

Suppose now that we fix $Q$ and $k$. 
Consider the dependence of the potential on $M$. 
Below some fixed value $M_{crit}$, 
$U$ possesses a single maximum; there is no well.
Above $M_{crit}$, $U$ possesses a well: 
with minimum $z_0$ (say), and maxima $z_-$ and $z_+$ on its left and 
its right respectively. We note also that $z_-$ (and never $z_+$) 
is always the absolute maximum of $U$. 
Lowering $m$ through $M_{crit}$ at a fixed values of $Q$ and $k$
we find that $z_0$ and $z_+$ (not $z_-$)  
coalesce when $m=M_{crit}$. 
The value $M_{crit}$ increases monotonically
from zero (as a function of both $Q$ and $k$).\footnote{We  remark that 
the technical details entering the determination of
boundaries such as $M_{crit}$ on the parameter plane have been 
discussed elsewhere by two of the authors 
in the context of global monopoles and will be omitted here. See 
Ref. \cite{IJ}. }

This completes the discussion of the topological form of 
the potential, as characterized by 
its critical points. This topology is not, however, 
always relevant physically. This will be the case if 
the well is not accessible physically.

The domains of $z$ which are physically accessible 
in the potential are determined by the mass shell condition 
Eq.(\ref{masss}). To locate these domains we examine 
where the critical points of the potential lie with 
respect to the fixed `energy' $-1$.
Again we fix $Q$ and $k$. These conditions will identify 
three values of $M$.

\item The upper limit on monotonic motion $M_{\cal M}$:

If $M$ is below some value $M_{\cal M}$ the absolute maximum of 
the potential will lie below 
the value $-1$. All values of $r$ are then accessible and all candidate 
physical trajectories necessarily monotonic --- either expanding from zero 
radius or collapsing to it.\footnote{When $M=M_{\cal M}$, $U(z_-)=-1$ and 
an unstable equilibrium with the wall poised precariously at 
$z_-$ is, of course, possible.} 
If $M>M_{\cal M}$ there are no monotonic trajectories. For each 
such $M$ there are always at least two trajectories, 
each with a single turning 
point, one bounded and another unbounded. When we refer to them below
we will describe the trajectory initially at rest at the turning 
points: the former collapses from a finite maximum to zero radius; 
the latter expands from a minimum to infinite radius. 

Whether the trajectories we have described translate into
configurations which are compatible with the boundary conditions 
is a question which we address in the following section. The Einstein
equations, as we will see, {\it do} 
admit spurious solutions which do not correspond to
the isolated lump of energy we are interested in.

The value $M_{\cal M}$ like $M_{crit}$ increases monotonically
from zero as a function of both $Q$ and $k$.

\item The limits of oscillatory motion, $M_0$ and $M_+$:

The analogue in our model of a radially 
deformed monopole corresponds to an oscillatory trajectory. 
These are  the only trajectories which should survive when gravity 
is turned off. When is such motion possible? 

To accomodate a stable oscillating trajectory in the potential,  
the well must be accessible on shell, $U(z_0) < -1$, and confine the motion 
on the right, $U(z_+)>-1$. Clearly these conditions will not 
be realized for every specification of $Q$ and $k$. When they are 
they will limit $M$ to values within a finite 
band $[M_0,M_+]$. The values $M_0$ at which $U(z_0) = -1$ and 
$M_+$ at which $U(z_+)= -1$ are indicated in Fig.~\ref{regmq}).
These two boundaries in the three dimensional parameter space coalesce 
on the boundary $M_{crit}$ along some critical curve $M_*$ where they 
terminate. For fixed $k$, we denote the limiting 
value of the charge on $M_*$ by
$Q_*(k)$. We have plotted $Q_*$ as a function of $k$ in 
Fig.~\ref{qtfinal}. Note that  $Q_*$ 
decreases monotonically to zero as $k\to \infty$. In particular,
the relationship $Q=Q_*(k)$ is invertible for the corresponding 
limiting value of $k$ at fixed $Q$, $k = k_*(Q)$.

Even without consulting spacetime diagrams, it
is already possible to conclude the following:

For a given $k$ there
exists a unique `static' trajectory for each $Q$ 
up a limiting value,
$Q_*(k)$; and that for a given $Q$, there is a corresponding limiting 
value of $k$, $k_*(Q)$. As we will see when we examine the 
corresponding spacetimes, not all `static' trajectories 
correspond to static spacetimes so that the physical limiting 
values will be lower.

There exists, at best, a finite spectrum $[M_0,M_+]$ 
bounded below by $M_0$, of stable oscillations about 
any static configuration. 

\end{enumerate}

Finally, we comment that the boundary structure on parameter 
space is captured completely by either of the two sections 
we have considered. The $M-Q$ section
contracts continuously
towards the unique fixed point, $M=0$, $Q=0$ as $k$ is raised. 
Its topological structure is unchanged.

In the following section, we will consider the embedding of the wall
trajectories in both the interior de Sitter and the exterior 
Reissner-Nordstr\"om spacetime.

\section{Embedding of the wall trajectory in spacetime}
\label{ebwm} 

In the present context, de Sitter space is represented most
conveniently by a Gibbons-Hawking diagram. For details, in the 
present context the reader is referred to Ref.\cite{bgg}. 
In this diagram the center is placed at the 
(north) pole of a round sphere. The evolution of this point is represented 
by the trajectory indicated $R=0$ on the left hand side of the spacetime 
diagram. The diagonal running from the upper right to the 
lower left represents the cosmological horizon of this point.

\begin{figure}
\begin{center}
\epsfig{file=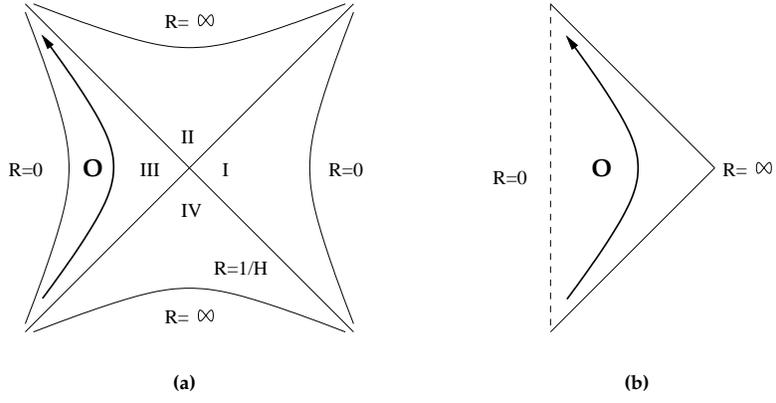,width=4in}
\vspace{1 cm}
\caption{
The trajectory type ${\cal O}$ occuring in the parameter regime ${\it S}$ 
embedded on (a) Gibbons-Hawking diagram for de Sitter spacetime;
(b) Penrose diagram for Reissner-Norsdtr\"om spacetime ($M<Q$)}
\label{rnwh}
\end{center}
\end{figure}

The core interior is represented by the spacetime region to the left of the 
trajectory on this diagram. It is clear that turning points of the motion of
the wall must occur within the static regions I and III with $R<H^{-1}$ where
the Killing vector $\partial_{T_S}$ is timelike, and $\partial_R$
spacelike. In particular, oscillatory solutions 
are necessarily confined to these regions
(one should not rule out, a priori, an oscillating core 
boundary in region I with an inflating interior). 
A globally static core must, however, lie in 
the left hand quadrant III. Any trajectory which crosses the horizon
necessarily inflates inside. See Fig.~\ref{rnwh}.

The nature of the 
Reissner-Nordstr\"om spacetime depends crucially on the charge to mass
ratio, $Q/M$. If $M<Q$ there are no horizons in the Reissner-Nordstr\"om 
geometry and a Penrose conformal diagram of its
maximal analytic extension consists of a single 
asymptotically flat globally static spacetime with a naked timelike
singularity at $R=0$. See Fig.\ref{rnwh}.
In our analysis, the Reissner-Nordstr\"om 
geometry will always be truncated at some
finite radius within which it is replaced by a patch of de Sitter space. 
If this radius does not fall to zero, the singularity does not 
appear in the physical spacetime and there is no physical justification to 
limit ourselves to values of $M$ exceeding $Q$ as one does in vacuum. 

\begin{figure}
\begin{center}
\epsfig{file=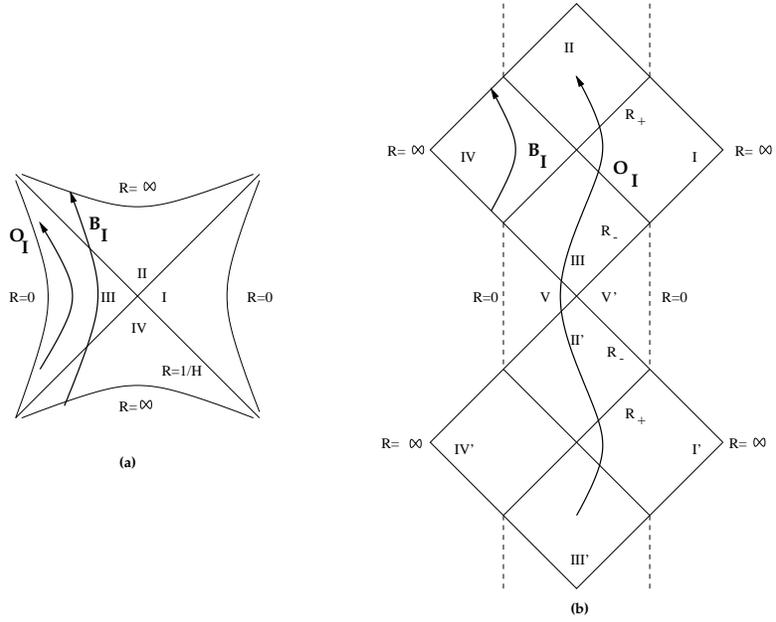,width=4in}
\vspace{1 cm}
\caption{
The trajectories ${\cal O}_I$ and ${\it B}_I$
occuring in parameter regime ${\it QSI}$ 
embedded on (a) Gibbons-Hawking diagram;
(b) Penrose-Carter diagram ($M>Q$).}
\label{rnh}
\end{center}
\end{figure}

The maximal analytic extension of the Reissner-Nordstr\"om geometry 
when $M>Q$ is represented on the Penrose-Carter(PC) diagram, Fig.~\ref{rnh}.
See Ref.\cite{carter} and also Ref.\cite{town} for a recent 
pedagogical discussion. This  consists of an infinite tower of 
identical connected universes. 
The singularities at $R=0$ are not visible at infinity in 
this geometry. Within the regions $R<R_-$ and 
$R>R_+$, the Killing vector $\partial_{T_M}$ is timelike: 
both of these regions are static. 
In the inter horizon region, $R_-<R<R_+$, 
$\partial_{T_M}$ becomes spacelike and $\partial_R$ generates 
temporal evolution.
The spacetime in this region is dynamical no matter how one 
cares to look at it. As in the interior de Sitter space, 
any turning points of the motion must occur in the static regions.
This will be useful to remember when locating 
turning points in spacetime.

We remark that the Cauchy horizon is unstable \cite{israel}. 
Under a small generic perturbation in the metric, it has been 
shown to collapse into a 
Schwarzschild type spacelike singularity limiting motion towards the future.
A consequence is that the exotic possibilities evoked by the 
Reissner-Nordstr\"om tower are irrelevant.
The life span of the physical system 
is limited to just one floor on this tower.

The exterior is represented by the spacetime region to the right of the 
trajectory on this spacetime diagram. The topology  
of a regular spatial slice is $R^3$ with a disk removed.

It can be shown that the fugacities $\beta_{D,M}$ are proportional to the  
derivative of the corresponding coordinate time with respect to proper 
time, 

\begin{equation}
\label{fugptime}
\beta_{D,M}=\mp A_{D,M} \dot{t}_{D,M}\,.
\end{equation}
In the case of de Sitter space and the $M>Q$ Reissner-Nordstr\"om geometry, 
the right hand side of Eq.(\ref{fugptime}) in turn relates 
these two functions to the course of the polar angle $\theta_{D,M}$
subtended by the trajectory about a fixed point in the corresponding 
spacetime diagram which permits the routing of the trajectory about this
point to be determined. For $\beta_D$ one has 

\begin{equation}
\beta_D \propto - \dot{\theta}_D\;,
\end{equation}
We note that $\beta_D>(<)0$ indicates clockwise (counter-clockwise)
motion about the origin.  

Unlike the de Sitter geometry, the Penrose-Carter diagram for
Reissner-Norsdtr\"om geometry with $M>Q$, possesses neither 
preferred origin, 
nor unique corresponding polar angle on the spacetime diagram. 
We consider the routing of the 
motion about the bifurcation points of
$R_+$ and $R_-$. We find

\begin{equation}
\beta_M \propto \mp  \dot{\theta^\pm}\;.
\end{equation}
where $\theta^{\pm}$ are the corresponding angles.

The interpretatinon of $\beta_M$ 
is different on the Reissner-Norsdtr\"om spacetime with $M<Q$.
Due to the absence of horizons, 
$\beta_M$ necessarily possesses a fixed sign. 
It is easily checked that $\beta_M$ must be positive for an isolated monopole 
with an infinite exterior. A negative $\beta_M$ in this 
case corresponds to a finite  
exterior geometry with a naked singularity. 

We are now in a position to describe the wall motion in spacetime which 
corresponds to any given set of parameters.

\section{Limit of stable oscillatory motion}
\label{bwtst}

We begin with a discussion  of trajectories which correspond to 
the intuitive notion of a monopole as a stable compact lump
of energy. As we have seen, such solutions must lie in 
the `oscillatory' regime of parameter space admitting bounded radial 
motion, with mass bounded below by $M_0$ and above by $M_+$. 
The boundary $M=Q$ provides a natural partition of this region. 
Indeed, we note that for low values 
of $Q$, $M_0 < Q$, with equality along  
$Q= Q_{0}(k)$. This value is strictly lower than $Q_*(k)$. 
The boundary $M_+$, on the other hand, lies strictly above 
$Q$ except along $Q= Q_{+}(k)$ where the two touch 
(with a common tangent). These two features are clearly indicated on the 
zoom-in of the $M - Q$ parameter plane. In Fig.~\ref{qtfinal} we 
plot $Q_{0}$ and $Q_{+}$ as functions of $k$. They clearly converge 
as $k$ becomes large. $Q_0$, $Q_+$  and $Q_*$
partition the oscillatory regime into three 
regions which we label {\it S, QSI}, and {\it QSII} on Fig.~\ref{regmq}.
The oscillatory motion compatible with each of these regions is different.

\noindent {\it S}: There exist stable oscillating trajectories with
both static interior and exterior:
in the interior, $\beta_D>0$ along the trajectory so that it lies in
region III of a Gibbons-Hawking diagram --- the interior does not inflate; 
the exterior Reissner-Nordstr\"om geometry with $M<Q$ is globally static,
The trajectory is indicated ${\cal O}$ in Fig.~\ref{rnwh}.

\noindent {\it QSI,II}: 
Stable oscillating trajectories would also appear to be admitted 
in these neighboring regions of parameter space.
However, whereas the interior geometry 
in both is essentially identical to that of an {\it S} trajectory,
the exterior geometry necessarily contains a black hole. 

If a genuine static trajectory with $M>Q$ exists, it must do so
along that section of the boundary $M_0$ where $M_0>Q$,
which occurs within {\it QSII}. Because $r$ is constant, it must lie 
entirely within one of the static regions with $R<R_-$ or $R>R_+$. 
Outside this domain, $R$ is a timelike coordinate and a
constant value of $R$ defines an impossible spacelike trajectory.

If $r<R_-$, the static trajectory lies within a 
black hole. Only if $r>R_+$ (with no horizon)
is  the exterior spacetime geometry 
globally static, so that we can speak of a genuinely 
static solution. There are, however, no solutions of this form:
within {\it QSII} the turning points $r_{min}$ and 
$r_{max}$ of oscillatory motion both 
lie below $R_-$.  In fact, the possibility 
$r_{min},r_{max}>R_+$, while consistent with the spacetime geometry, 
never occurs. Stable static monopoles 
(and stable radial oscillations about them), 
appear always to configure themselves so that $M<Q$. 

For a given $Q$ there exists an upper bound on $\eta$
admitting such a solution 
(as for a given $\eta$ there exists an upper bound on  
$Q$), determined by the crossing of $M_0$ and $M=Q$,
strictly below the `naive' bound on the `oscillatory' regime 
discussed in Sec.~\ref{ssfK}. The existence of the limit on $\eta$ 
was predicted within the 
simplified zero surface tension model by 
Tachizawa, Maeda and Torii in Ref.~\cite{tachi}. The existence of this limit 
was also noted by Sakai in \cite{sk}. 
This value of $\eta$ signals the onset of topological inflation. 

Consider, now, the fate of an oscillating trajectory
as $M$ is raised through $M_{hor}$
from some initial value in {\it S} maintaining $Q$ and $k$
constant. Because $M_{hor}>M_0$ in this regime, 
the trajectory must undergo finite oscillation in $r$
(there are no static trajectories). The 
surplus $M$ provides the wall with radial kinetic energy.
As $M_{hor}$ is crossed, two horizons with initially 
equal radii appear in the 
exterior geometry. Where the turning points of the 
oscillatory motion, $r_{min}$ and $r_{max}$ say, lie with respect to the 
horizons at $R_+$ and $R_-$ will depend on the values of $Q$ 
and $k$. 

If $Q<Q_+(k)$,  {\it QSI} is entered with $r_{min} <R_-$ 
and $r_{max}> R_+$; whereas if $Q$ lies between $Q_+(k)$ and 
$Q_0(k)$, {\it QSII} is entered with both $r_{min}$ and 
$r_{max}$  less than $R_-$.\footnote{We note that there 
is also a region within {\it QSII} corresponding to 
values of $Q$ and $k$ within the range $[Q_0(k),Q_*(k)]$ which 
cannot be considered as excitations of any {\it S} static configuration.}
When $Q=Q_+$, $r_{max}$ coincides with the right maximum of the 
potential and $r_{max}=R_-=R_+$.

\begin{figure}
\begin{center}
\epsfig{file=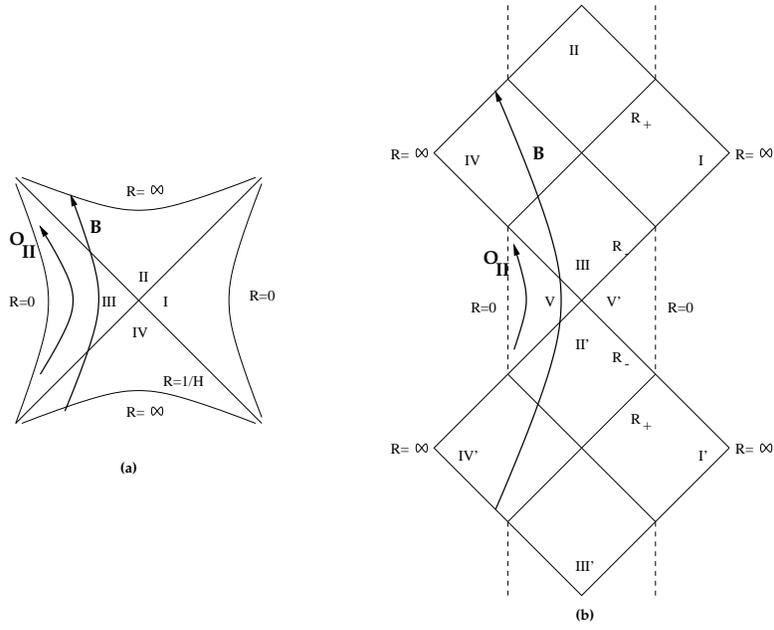,width=4in}
\vspace{1 cm}
\caption{
A trajectory of type ${\cal O}_{II}$ 
occuring in parameter regime ${\it QSII}$, and
a trajectory of type ${\cal B}$ 
occuring in parameter regime ${\it B}$ (which includes ${\it QSII}$)
embedded on (a) Gibbons-Hawking diagram;
(b) Penrose-Carter diagram.}
\label{rnh2}
\end{center}
\end{figure}

The exterior spacetimes which correspond to 
`oscillatory' trajectories ${\cal O}_I$ 
in {\it QSI} and ${\cal O}_{II}$  in 
{\it QSII} are illustrated in 
Figs.\ref{rnh} and \ref{rnh2}, respectively.

The `oscillatory' trajectory ${\cal O}_I$ 
interpolates between a maximum in region I and a minimum in  region 
$V$. Its apparent subsequent oscillatory course up through the 
Penrose-Carter tower is an analytical accident without 
any observable consequences. The physical solution clearly does not 
oscillate coming as it will to the unpleasant end described in the 
previous section as it crosses 
the Cauchy horizon. The exterior geometry collapses in a black hole. 

The trajectory ${\cal O}_{II}$ oscillate within region $V$. The exterior
geometry again collapses in a black hole isolating the monopole inside. 
The gauge monopole analogues
of both solutions were observed numerically by Sakai in \cite{sk}. 
Their zero tension analogue was identified in 
Ref.~\cite{tachi} by Tachizawa, Maeda and Torii.

Finally, we note that the Penrose singularity theorem places no classical 
obstruction on the formation of any of the solutions we have described 
from non singular initial conditions, be they static or 
black hole.\cite{fg,fgg}

\section{ Lower bound on the monopole mass}
\label{lb}

The oscillatory solution in {\it S} described above is the 
only solution of the Einstein equations satisfying the 
boundary conditions which corresponds to an isolated monopole in the 
parameter regime $M<Q$. Technically, this is because 
$\beta_M<0$ along the remaining trajectories,
be they monotonic, collapsing or expanding bounces. 
This is just as well: 
while gravity might be sufficiently strong to 
provoke the collapse of a charged object, one would not expect this 
to happen if the charge exceeds $M$;  
nor would one expect gravity 
to promote the explosion of a monopole. 
A negative $\beta_M$ corresponds, in the regime $M<Q$,
to an exterior which is a finite region of the 
Reissner-Nordstr\"om geometry with an unphysical naked  singularity at 
the antipode. The spatial geometry is a closed three sphere which is 
inconsistent with the asymptotically flat boundary conditions that 
we associate  with an isolated monopole. 
Because it contains a naked singularity we  
consider it unphysical. 

An immediate corollary of the above observation is 
the existence of a lower bound on the mass of 
a physically realistic configuration, static or otherwise:
(i) if $Q<Q_0(k)$, so that  stable static solutions exist,
this value is $M_0$\cite{SW};
(ii) if $Q\ge Q_0(k)$ 
and there do not, a bound is provided by $Q$.
The later bound will be sharpened below.

For a constant $k$, the mass of a static
solution $M_0 \simeq Q ({Q/Q_0})^{1/2}$ which has the 
same functional form as the Minkowski space limit.

\section{Inflating Bounces with $M> Q$}  
\label{bouncy}

In Sec.~\ref{ssfK}, bounce solutions were 
identified in the parameter regime bounded below by $M_{\cal M}$. 
Such solutions coexist with the quasi-static solutions we have 
described in each of {\it QSI} and {\it QSII}. 

We again discard the 
collapsing solution as an unphysical closed universe with a naked 
singularity. However, the expanding bounce trajectories are 
consistent with the boundary conditions.

The regime admitting bounces partitions naturally into
three regions: {\it QSI} (as before), ${\it B}$ (which contains
{\it QSII}) indicated on Fig.~\ref{regmq},
and ${\it B}'$ indicated on Fig.~\ref{regmqzi}.

\begin{figure}
\begin{center}
\epsfig{file=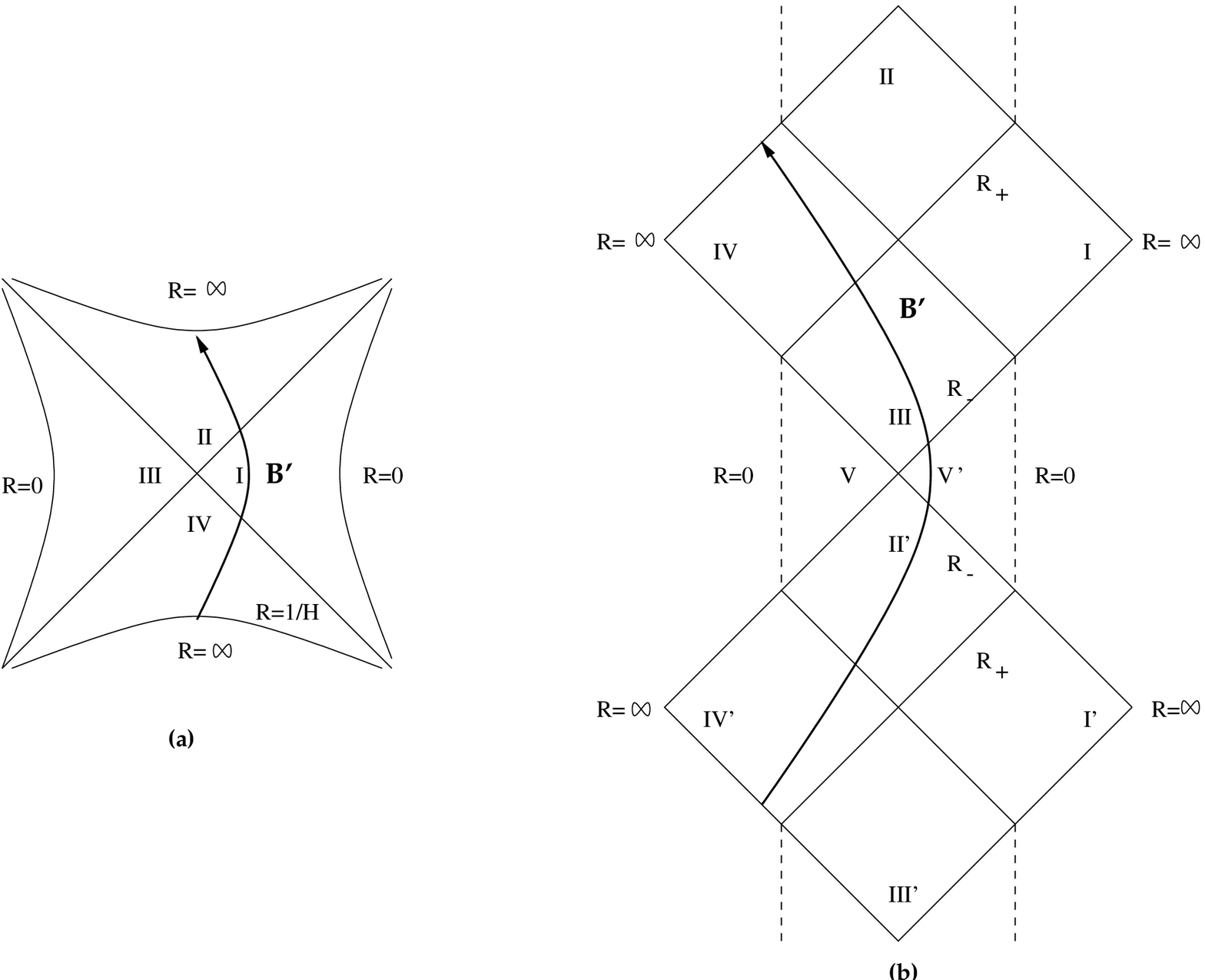,width=4in}
\vspace{1 cm}
\caption{
A trajectory of type ${\cal B}'$ occuring in parameter regime ${\it B}'$
embedded on (a) Gibbons-Hawking diagram;
(b) Penrose-Carter diagram.}
\label{rnh3}
\end{center}
\end{figure}

The interior spacetime of an expanding bounce clearly inflates. 
The trajectories are embedded on the Reissner-Nordstr\"om spacetime
as ${\cal B}_I$ on Fig.~\ref{rnh} for {\it QSI};
${\cal B}$ on Fig.~\ref{rnh2} for {\it B};
and ${\cal B}'$ on Fig.~\ref{rnh3}.  for ${\it B}'$.

The expansion in all cases occurs behind a throat geometry which 
subsequently collapses into a Reissner-Nordstr\"om black hole
(in the same way as it does outside the oscillatory counterparts
discussed in Sec.~\ref{bwtst})  
This expansion does not occur at the expense of the exterior geometry 
but (with respect to a reasonable slicing
of spacetime) does get cut off from the exterior by the formation of 
a black hole.

Qualitatively, the bounce occuring in {\it QSI} is 
very similar to the
false vacuum bubble bounces  described in Ref.\cite{bgg}.
Note that the Penrose singularity theorem places an obstruction
to its formation from non singular initial conditions. 
Accessible or not classically, this
trajectory is of interest because of the 
possibility of tunneling into it from its bounded counterpart, \cite{fgg}.

The bounce occuring in {\it B} is very different, 
contracting from infinity in one asymptotically
flat region of the Reissner-Nordstr\"om spacetime
to a minimum on the left hand side of the Penrose-Carter tower
before expanding to the 
corresponding asymptotically flat region  on the next floor 
of the tower. Clearly, the full bounce is not a 
physically realizable configuration. The physically relevant leg of any 
bounce is its expansion from a stationary minimum. Bounces
correspond either to the thermodynamical or quantum mechanical 
materialization of a configuration.
 
Interestingly, there is a narrow window in the
neighborhood of this stationary initial configuration 
where the Penrose theorem does not present any
obstruction to the classical assembly of the {\it B} bounce from 
non-singular initial conditions. These are the analogues of 
topologically inflating gauge monopoles.

In his numerical simulations of the dynamics
of gauge monopoles, Sakai also
observed inflating monopoles (corresponding to our bounces)
to co-exist with collapsing monoples (corresponding 
our ${\cal O}_I$). 

Bounces occuring in the narrow regime of parameter space indicated
${\it B}'$ occur in a convex potential. Whereas the  asymptotics of 
such bounces are identical to those for {\it B}, 
their minimum occurs now on the right of the Penrose-Carter tower. 
In contrast to {\it B} bounces, the Penrose theorem implies 
that its formation is unphysical 
on the complete expanding leg. On its 
contracting leg, there is no asymptotically flat spatial slice
containing the trajectory. We must conclude that such trajectories
are unphysical.

\section{All Monotonic trajectories are unphysical}  
\label{mono}

\begin{figure}
\begin{center}
\epsfig{file=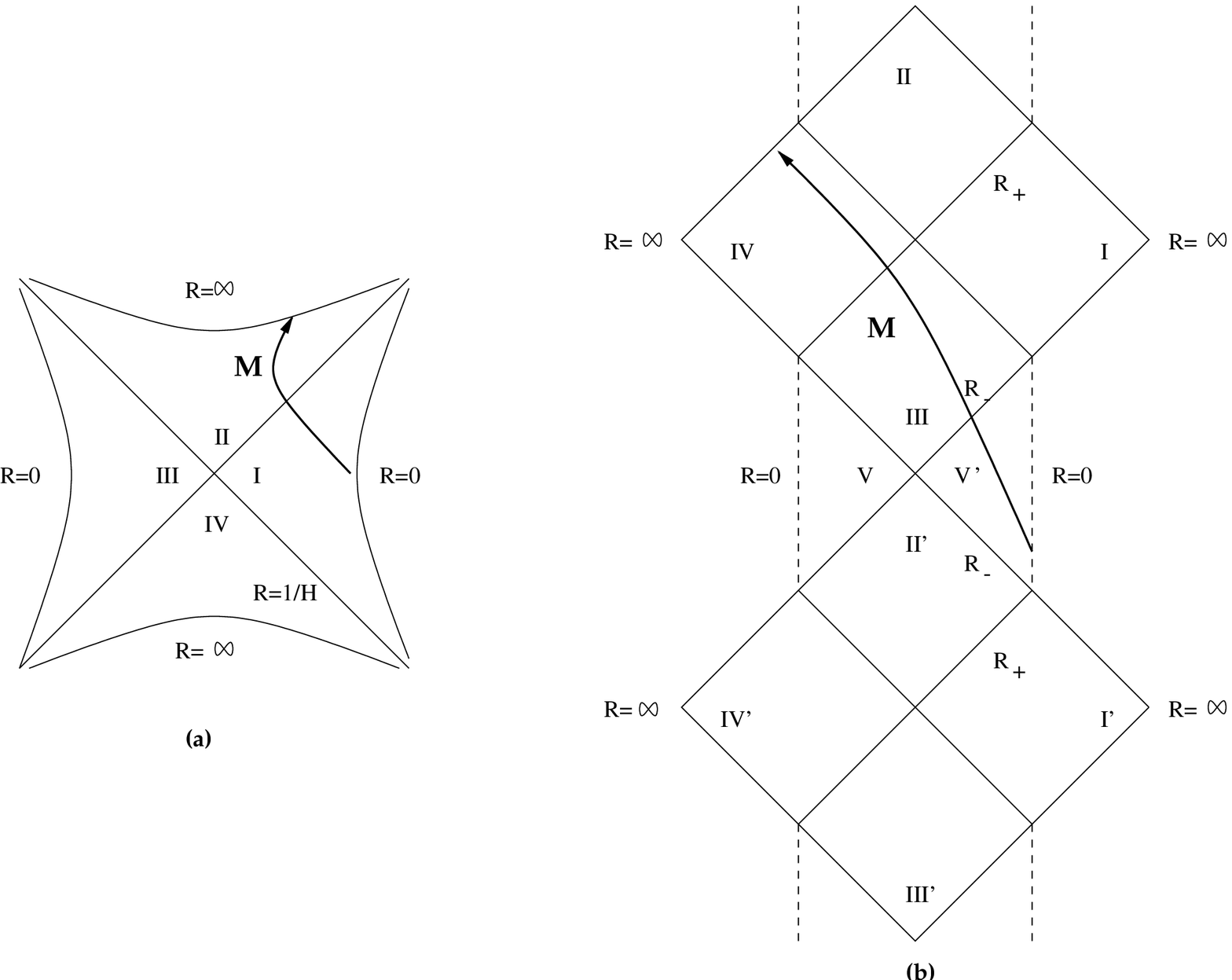,width=4in}
\vspace{1 cm}
\caption{
A trajectory of type ${\cal M}$ occuring in parameter regime ${\it M}$
embedded on (a) Gibbons-Hawking diagram;
(b) Penrose-Carter diagram ($M>Q$).}
\label{rnh4}
\end{center}
\end{figure}

We have already discounted monotonic trajectories with $M<Q$.
The boundary $M_{crit}$ partitions the remainder of this  regime.
In the bounded regime $M_{\cal M} < M_{crit}$ the potential is convex
and both $\beta_D$ and $\beta_M$ possess definite signs.
We have indicated the trajectory by ${\cal M}$ on Fig. \ref{rnh4}. 
In the remaining unbounded region with $M_{crit}< M_{\cal M}$, 
the effective potential develops a well and  both $\beta_D$ and 
$\beta_M$  change sign in the course of their evolution.
Apart from this single dynamical detail,
motion is qualitatively identical in both regimes.

Is this motion physical from a classical point of 
view? The part of the trajectory lying within $r<R_-$ 
necessarily contains a naked singularity in its exterior;
moreover, the interior initially contains a three-sphere's 
worth of de Sitter
space. The trajectory is clearly unphysical in this regime. 
In fact, the Penrose theorem  
forbids the assembly of such a 
trajectory by classical means\cite{fgg}. We dismiss this solution as 
unphysical. It would appear that there are no physical
monotonic trajectories in this model. 

If we take in account the elimination as unphysical 
of all possible trajectories in both 
${\cal M}$ and ${\it B}'$, the lower bound on the mass of any 
asymptotically flat configuration is raised. As $Q$ increases 
above $Q_0$, the lower bound follows the line
$M=Q$, then $M=M_{crit}$, and finally $M=M_{\cal M}$.

\section{False vacuum bubble limit}
\label{BGG}

We are finally in a position to 
examine the limit $Q\to 0$, 
where the model had better reproduce the ``false vacuum
bubble'' investigated by Blau, Guendelman and Guth and others. 
Briefly, for each value of $M$ below some critical mass 
$M_{cr}$, both collapsing and expanding bounce motion occur
as we described in our introduction.
For masses above $M_{cr}$ all 
motion is monotonic: the core expands from a singular zero radius
behind the Schwarzschild horizon and like the 
bounce described in the introduction is connected to the 
asymptotically flat region by a throat. Though the throat collapses, 
the core expands forever with an inflating interior. 
The reader is referred 
to \cite{bgg} for details. 
Both the expanding bounce and the monotonic solution violate the 
Penrose theorem along their course \cite{fg}. 

This limit should be consistent with solutions lying on the $M$-axis on the 
$M-Q$ section. At first sight, however,
the limit $Q\to 0$ of our model appears to contradict their findings. 
Specifically, there does not appear to be
any analog of $M_{cr}$ at $Q=0$ --- the monotonic trajectories we
find do not even exist in this regime. 	
To resolve this apparent contradiction,
note that, as $Q\to 0$ in this regime, 
the left maximum of the potential $U$
occurs at ever decreasing radius ($z_-\to 0$) while, simultaneously, 
the well depth becomes infinitely deep, $U(z_0)\to -\infty$. 
We also note that, as $Q\to 0$, 
the inner horizon of the Reissner-Nordstr\"om geometry
appoaches zero, $R_-\to 0$, while 
the outer horizon at $R=R_+$ becomes the Schwarzchild horizon.
The bounce trajectory occuring in ${\cal B}$ thus approaches arbitrarily 
close to $r=0$, the Penrose window we discussed in Sec.~\ref{bouncy}
closes and the trajectory on its expanding leg becomes indistinguishable 
from a monotonically growing false vacuum bubble. 
It is clear that we should identify $M_{cr}$ with $M_+$ at $Q=0$, 
not with $M_{\cal M}$. 

We also note that below $M_+$, in {\it QSI} the 
a quasi-oscillatory trajectory 
approaches $z=0$ arbitrarily closely and become indistinguishable from 
a collapsing bounce. The expanding bounce, as we commented earlier
does not suffer any signifacant local change.

\section{Concluding Remarks}
\label{con}

We have examined in some detail the dynamics of a charged false vacuum
bubble within the thin wall approximation. 
We claim that, with the identification of $Q$ with the 
magnetic charge $g$ (related to the electric charge by $g=4\pi/e$)
the model mimics the radial dynamics of a spherically symmetric
magnetic monopole. In particular, the model provides a
valuable guide to understanding the physics which 
underlies both the onset of instability of a static monopole
as well as the conditions which need to be met to
produce a topologically inflating object.

It would appear that inflation does not necessarily require 
dialling up the symmetry-breaking scale $\eta$; 
an inflating solution only requires that 
the ADM mass be sufficiently large, which is possible in principle
for arbitrarily low values of $\eta$ or $Q$. 
In this respect, the monopole we consider differs from the
`global' monopole discussed in \cite{IJ} where inflation is only 
possible when $\eta$ is raised above the Planck scale. 
However, it should also be pointed out that 
in a field theory of monopoles the mass is itself a function 
of $\eta$. It is not clear if the high mass and low $\eta$
inflating solutions we find can be realized in practice.

There are a few interesting extensions of this work:

We note that, for every monopole which collapses into a 
black hole in the parameter regimes 
{\it QSI} and {\it QSII}, there will be a corresponding 
expanding bounce configuration 
with identical values of the conserved mass and charge. 
On semiclassical grounds, one would anticipate 
a finite amplitude for tunneling from the former to the latter. 
The construction of the instanton mediating this passage
should provide a valuable exercise in semi-classical 
quantum gravity.

We have considered a description of a spherically symmetric
field theoretical monopole 
in which its core boundary is modeled as 
a relativistic membrane. How robust is this description when 
spherical symmetry is relaxed?

In the seventies it was shown that, 
in Dirac's extensible model of the 
electron, the static charged membrane 
is unstable to non-radial deformations\cite{Kuti}. 
The origin of this instability is similar to that which triggers 
fission of the atomic nucleus (the boundary conditions 
differ). On the other hand, 
in Ref.\cite{Gold} Goldhaber argued that a global monopole 
suffers from a cylindrical string-like instability. Superficially,
this would appear to be analogous to the spike (zero area) instability 
of a Nambu-Goto membrane. However, it is 
likely that higher curvature (rigidity)
corrections to the Nambu-Goto action must be included 
to model the field theory when spherical symmetry is relaxed.
Such additions would tend to moderate (or eliminate) 
the instabilities of Nambu-Goto membranes. It would be interesting to 
explore the membrane-topological defect correspondence 
in greater detail.

\acknowledgments

Thanks to Gilberto Tavares for technical assistence. G.A. was supported 
by a CONACyT graduate fellowship. I.C. was supported 
in part by the Institute of Cosmology at Tufts University.  
The work of J.G. has received support from  
DGAPA at UNAM, CONACYT proyect 32307E and a CONACyT-NSF collaboration.

\end{multicols}

\newpage

\newpage

\newpage

\newpage

\newpage

\newpage

\newpage

\newpage

\newpage

\begin{references}

\bibitem{Linde} A. Linde, Phys. Lett. {\bf B327}, 208 (1994).

\bibitem{Alex} A. Vilenkin, Phys. Rev. Lett. {\bf 72}, 3137 (1994).

\bibitem{Eduardo} E. Guendelman and A. Rabinowitz, Phys. Rev. {\bf D44}, 
3152 (1991).

\bibitem{bgg} S. K. Blau, E. I. Guendelman and A. H. Guth, 
Phys. Rev. {\bf D35}, 1747 (1987).

\bibitem{ag} Alan Guth, {\it The Inflationary Universe} (Addison-Wesley,
Reading, MA. 1997).

\bibitem{bv} M. Barriola and A. Vilenkin, Phys. Rev. 
Lett. {\bf 63}, 341 (1989).

\bibitem{hl} D. Harari and C. Lousto, Phys. Rev. {\bf D42}, 2626 (1990).  

\bibitem{gibb} G. W. Gibbons, in {\it Proceedings of the XII
Autum School on the Physical Universe, Lisbon, 1990}, edited
by J.D. Barrow {\it et al.}, Lecture Notes
in Physics Vol. 383 (Springer Verlag, Berlin, 1991).

\bibitem{ortiz} M. E. Ortiz, Phys. Rev. {\bf D45}, R2586 (1992).

\bibitem{bfm} P. Breitenlohner, P. Forgacs and D. Maison, Nucl. Phys.
{\bf B383}, 357 (1992); {\it ibid}. {\bf B442}, 126 (1995).


 \bibitem{sstm} N. Sakai, H. A. Shinkai, T. Tachizawa, and K. Maeda,
Phys. Rev. {\bf D53}, 655 (1996); {\it ibid}. {\bf D54}, 2981 (1996).
 
\bibitem{sk} N. Sakai, Phys. Rev. {\bf D54}, 1548 (1996).

\bibitem{IJ} I. Cho and J. Guven, Phys. Rev. {\bf D58}, 63502 (1998).


\bibitem{Dirac} P. A. M. Dirac, Proc. R. Soc. {\bf A 268}, 57 (1962).

\bibitem{tachi} T. Tachizawa, K. Maeda and T. Torii, 
Phys. Rev. {\bf D51}, 4054 (1995).

\bibitem{lee} K. Lee, V. P. Nair and E. Weinberg, 
Phys. Rev. {\bf D45},  2751 (1992).

\bibitem{trodden} G. L. Alberghi, D. A. Lowe and M. Trodden, 
JHEP 9902:020 (1999).

\bibitem{CG} For a rigorous justification of
this approximation in the context of a domain wall,
see B. Carter and R. Gregory, Phys. Rev. {\bf D51}, 5839 (1995).

\bibitem{carter} B. Carter, Phys. Lett. {\bf 21}, 423 (1966).


\bibitem{town} P. K. Townsend, {\it Black Holes}, 
Lecture Notes, gr-qc/9707012. 


\bibitem{israel} 
See, for example, W. Israel, in  {\it `Black Holes,
Classical and Quantum'} (Mazatl\'an, Mexico, 1998).

\bibitem{fg} E. Farhi and A. H. Guth, Phys. Lett. 
{\bf B183}, 149 (1987).

\bibitem{fgg} E. Farhi, A. Guth and J. Guven, Nucl. Phys. {\bf B339}, 
417 (1990).

\bibitem{SW} This is consistent with the general results of
D. Sudarsky and R. Wald, Phys. Rev. {\bf D47}, 5209 (1993); 
{\it ibid}. {\bf D46}, 1453 (1992). 



\bibitem{Kuti} P. Hasenfratz and J. Kuti, 
Phys. Rep. {\bf 40}, 75 (1978);
P. Gnadig, Z. Kunszt, P. Hasenfratz, and J. Kuti, 
Annals of Phys. {\bf 116}, 380 (1978).

\bibitem{Gold} A. Goldhaber, Phys. Rev. Lett. {\bf 63}, 2158 (1989).

\end{references}
\end{document}